\documentclass{evn2004}
\usepackage{txfonts}
\usepackage{graphicx}
\begin{document}
   \title{Extending and Exploring the 2\,cm Survey Sample}

   \author{Eduardo Ros}

   \institute{Max-Planck-Institut f\"ur Radioastronomie, Auf dem H\"ugel 69, D-53121 Bonn, Germany
             }

   \abstract{
We present new results from the VLBA 2\,cm Survey, an imaging survey of
active galactic nuclei (AGN) at sub-milliarcsecond resolution. We extend
the structural variability monitoring program of more than 130
extragalactic parsec-scale radio jets from over 170 AGN to a total of eight
years.  The sample is explored further in time for all sources, and
intensively for individual objects.  We report new detailed results on
the compact sources III\,Zw\,2, AO\,0235+16, and NRAO\,512.
   }

   \maketitle

%
\section{Introduction}

Over the last decade we have used the Very Long Baseline 
Array
(VLBA) at 15\,GHz to image the structure of more than 170 active
galactic nuclei (AGN).  This large collaboration is known as the ``2\,cm Survey".
Most of the sources show a core-jet morphology with typical jet 
structures (e.g., Zensus \cite{zen97}).  
The 2\,cm Survey aims to study the jet kinematics in AGN to better
understand the processes involved in the formation and evolution
of powerful jets.
The first images of 132 sources, selected from data taken
between August 1994 and October 1996, were
presented by Kellermann et al.\ (\cite{kel98}).  An additional
39 sources were presented by Zensus et al.\ (\cite{zen02}), from
data collected from August 1997 to March 2001.
Many of the jets may be described by a small number of apparently 
discrete features, which are generally called components.
Kellermann et al.\ (\cite{kel04}) reported the sky motions over
the above mentioned time lapse
of 208 components in 110 AGN from the sample.
We have made all the survey images available
on the Internet
under \verb+http://www.nrao.edu/2cmsurvey+.

We continued adding new observations
from October 2001 to October 2002.  We also included data
from other survey projects
in our kinematic analysis,
such as the pre-VSOP observations from L.\ I.\ Gurvits et al.\ 
(in prep.), the S5 Polar Cap Sample observations
by P\'erez-Torres et al.\ (\cite{per04}), and the 
observations from GHz-peaked-spectrum radio sources by 
Stanghellini et al.\ (\cite{sta01}).
To continue beyond the observations reported here, we have
assembled a complete flux-density limited sub-sample from the full
2\,cm VLBA survey.  The selection criteria are described in
Kellermann et al.\ (\cite{kel04}).  Observations
of the complete sample are being continued since May 2001, under the name of
``\underline{Mo}nitoring of 
\underline{J}ets in \underline{A}GN with 
\underline{V}LBA \underline{E}xperiments (MOJAVE)", see
\verb+http://www.physics.purdue.edu/astro/MOJAVE/+.
This program includes also dual polarisation observations.

In this contribution we present results based on the 2001--2002 
observations, which turn out to be complementary to 
the previously published ones.

%
\section{Kinematics\label{sec:kinematics}}

\begin{figure*}[tbh!]
\begin{center}
\includegraphics[clip,width=0.9\textwidth]{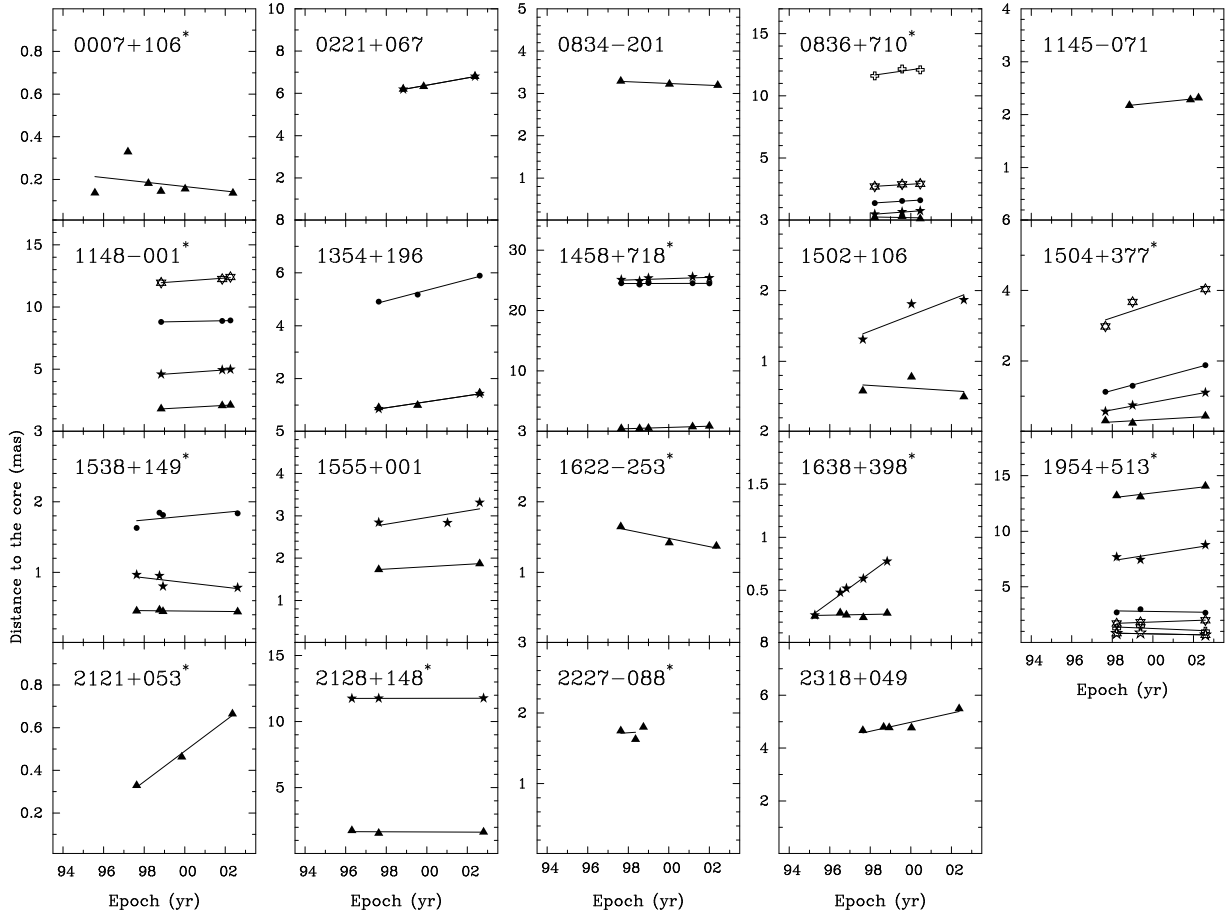}
\end{center}
\vspace{-10pt}
\caption{Plots showing the change in separation with time of features
of one-sided jets for which we have measured a velocity from observations
at three or more epochs, after including the data from
the newer observations and from another 15\,GHz surveys (see text).
The codes used are the same as in Fig.~1 of
Kellermann et al.\
(\cite{kel04}).  An asterisk denotes sources that were model fitted
in the $(u,v)$-plane rather than in the image plane.  The solid lines
denote the best least-squares linear fit to the data, and the slope
represents the proper motion, $\mu$.
\label{fig:kinematics}
}
\end{figure*}

The survey observations reported so far (Kellermann et al.\ \cite{kel98};
Zensus et al.\ \cite{zen02}; Kellermann et al.\ \cite{kel04})
included 29 observing sessions.  We added 18
observing sessions with typically eight sources per session, where
each source was observed once per hour for 6\,min to 
8\,min over a range of 8\,h
in hour angle.  The data reduction procedures for imaging,
component identification and measurement of its
position, combining {\sc difmap}
and $\mathcal{AIPS}$ tasks, are described in Kellermann et al.\ 
(\cite{kel98}; \cite{kel04}).
Our additional data allowed us to analyse the kinematics of 41 components
in 19 
sources.  The measured positions and speeds
are presented in Fig.\ \ref{fig:kinematics}
and tabulated column 5 of Table~\ref{table:kinematics}. 
In some cases, when the jet features were not well resolved, or in
other peculiar cases, we performed a fit to the visibilities
using Gaussian functions.
In this case, we have marked an asterisk at the 
name of the sources in Fig.\ \ref{fig:kinematics}.

\begin{table}[t]
\caption{Source Kinematics\label{table:kinematics}}
\begin{flushleft}
\[
\centering
\resizebox{\columnwidth}{!}{%
\begin{tabular}{@{}l@{\,}c@{\,}ccccc@{}}
\hline
\hline
\noalign{\smallskip}
           & Optical & \\
Object     & class$^\mathrm{a}$ & Id & N & $\mu$ & $\beta_{\rm app}$$^\mathrm{b}$ & $t_0$$^\mathrm{c}$ \\
      &           &       &   & {\scriptsize [mas\,yr$^{-1}$]} & & {\scriptsize [yr]} \\
\noalign{\smallskip}
\hline
\noalign{\smallskip}
0007$+$106 & G & B & 6 & $-0.011\pm  0.015$ & $-0.06\pm  0.09$ & $\dots$ \\
0221$+$067 & Q & B & 3 & $ 0.173\pm  0.013$ & $ 5.26\pm  0.41$ & 1963.0 \\
0834$-$201 & Q & B & 3 & $-0.021\pm  0.005$ & $-2.0\pm  0.5$ & $\dots$ \\
0836$+$710$^\mathrm{d}$ 
           & Q & B & 3 & $-0.03\pm  0.08$ & $-2.6\pm  6.7$ & $\dots$ \\
           &   & C & 3 & $ 0.118\pm  0.017$ & $10.1\pm  1.4$ & 1994.2 \\
           &   & D & 3 & $ 0.105\pm  0.021$ & $ 9.0\pm  1.8$ & 1985.1 \\
           &   & E & 3 & $ 0.096\pm  0.022$ & $ 8.3\pm  1.9$ & 1970.0 \\
1145$-$071 & Q & B & 3 & $ 0.039\pm  0.005$ & $ 2.52\pm  0.34$ & $\dots$ \\
1148$-$001 & Q & B & 3 & $ 0.089\pm  0.004$ & $ 7.23\pm  0.32$ & 1978.5 \\
           &   & C & 3 & $ 0.115\pm  0.003$ & $ 9.32\pm  0.24$ & 1958.8 \\
           &   & D & 3 & $ 0.031\pm  0.007$ & $ 2.53\pm  0.60$ & $\dots$ \\
           &   & E & 3 & $ 0.118\pm  0.029$ & $ 9.7\pm  2.3$ & $\dots$ \\
1354$+$196$^\mathrm{e}$
           & Q & B & 3 & $ 0.115\pm  0.028$ & $ 4.7\pm  1.2$ & 1990.2 \\
           &   & D & 3 & $ 0.202\pm  0.025$ & $ 8.2\pm  1.0$ & 1973.5 \\
1458$+$718$^\mathrm{e}$ 
           & Q & B & 5 & $ 0.102\pm  0.012$ & $ 4.9\pm  0.6$ & 1993.8 \\
           &   & C & 5 & $ 0.11\pm  0.06$ & $ 5.3\pm  2.9$ & $\dots$ \\
           &   & D & 5 & $ 0.010\pm  0.033$ & $ 0.5\pm  1.6$ & $\dots$ \\
1502$+$106 & Q & B & 3 & $-0.02\pm  0.06$ & $-1.4\pm  4.3$ & $\dots$ \\
           &   & C & 3 & $ 0.11\pm  0.05$ & $ 8.7\pm  4.2$ & 1985.1 \\
1504$+$377$^\mathrm{e}$
           & G & B & 3 & $ 0.032\pm  0.023$ & $ 1.2\pm  0.9$ & 1989.6 \\
           &   & C & 3 & $ 0.107\pm  0.006$ & $ 4.11\pm  0.25$ & 1992.2 \\
           &   & D & 3 & $ 0.156\pm  0.008$ & $ 5.98\pm  0.30$ & 1990.5 \\
           &   & E & 3 & $ 0.19\pm  0.09$ & $ 7.3\pm  3.4$ & 1981.1 \\
1538$+$149 & Q & B & 4 & $-0.003\pm  0.003$ & $-0.11\pm  0.10$ & $\dots$ \\
           &   & C & 4 & $-0.033\pm  0.021$ & $-1.2\pm  0.7$ & $\dots$ \\
           &   & D & 4 & $ 0.027\pm  0.027$ & $ 0.97\pm  0.94$ & $\dots$ \\
1555$+$001 & Q & C & 3 & $ 0.08\pm  0.07$ & $ 6.0\pm  5.6$ & $\dots$ \\
1622$-$253 & Q & B & 3 & $-0.059\pm  0.022$ & $-2.5\pm  0.9$ & $\dots$ \\
1638$+$398 & Q & B & 5 & $ 0.003\pm  0.008$ & $ 0.24\pm  0.57$ & $\dots$ \\
           &   & C & 5 & $ 0.139\pm  0.008$ & $10.2\pm  0.6$ & $\dots$ \\
1954$+$513 & Q & B & 3 & $ 0.22\pm  0.09$ & $13.2\pm  5.1$ & 1993.2 \\
           &   & C & 3 & $ 0.28\pm  0.13$ & $17.1\pm  7.9$ & $\dots$ \\
           &   & D & 3 & $-0.03\pm  0.07$ & $-1.6\pm  4.3$ & $\dots$ \\
           &   & E & 3 & $ 0.065\pm  0.009$ & $ 3.9\pm  0.6$ & $\dots$ \\
           &   & F & 3 & $-0.08\pm  0.07$ & $-4.5\pm  4.5$ & $\dots$ \\
           &   & G & 3 & $-0.036\pm  0.019$ & $-2.19\pm  1.17$ & $\dots$ \\
2121$+$053 & Q & B & 3 & $ 0.071\pm  0.006$ & $ 5.64\pm  0.46$ & 1993.2 \\
2128$+$048$^\mathrm{f}$ 
           & G & B & 3 & $-0.007\pm  0.030$ & $-0.37\pm  1.55$ & $\dots$ \\
           &   & C & 3 & $ 0.003\pm  0.001$ & $ 0.16\pm  0.04$ & $\dots$ \\
2227$-$088$^\mathrm{e}$ 
           & Q & B & 3 & $ 0.02\pm  0.16$ & $ 1.3\pm 11.1$ & $\dots$ \\
2318$+$049$^\mathrm{e}$ 
           & Q & B & 5 & $ 0.170\pm  0.042$ & $ 6.1\pm  1.5$ & 1970.7 \\
\noalign{\smallskip}
\hline
\end{tabular}
}
\]
\begin{scriptsize}
\begin{list}{}{
\setlength{\leftmargin}{6pt}
\setlength{\rightmargin}{0pt}
}
\item[$^{\mathrm{a}}$] Optical classification according to the 
V\'eron-Cetty \& V\'eron (\cite{ver01}) catalog, where Q\,=\,quasar
and G\,=\,galaxy.
\item[$^{\mathrm{b}}$] To compute the linear speed, and throughout
the paper we use a cosmology with $H_0=70$\,km\,s$^{-1}$\,Mpc$^{-1}$,
$\Omega_m=0.3$, and $\Omega_\Lambda=0.7$.
\item[$^{\mathrm{c}}$] Extrapolated epoch of origin.
\item[$^{\mathrm{d}}$] Includes data from the S5 
Polar Cap Sample (P\'erez-Torres et al.\ \cite{per04}).
\item[$^{\mathrm{e}}$] Includes data from the pre-VSOP observations
by L.\ I.\ Gurvits et al.\ (in prep.).
\item[$^{\mathrm{f}}$] Includes data from observations taken
in April 1996 on Gigahertz-peaked sources and published 
by Stanghellini et al.\ (\cite{sta01}).
\end{list}
\end{scriptsize}
\end{flushleft}
\end{table}

\begin{table}[t]
\caption{Sources without reported kinematics\label{table:nokin}}
\begin{flushleft}
\[
\centering
\resizebox{\columnwidth}{!}{%
\begin{tabular}{@{}l@{\,}c@{\,}clc@{}}
\hline
\hline
\noalign{\smallskip}
 & Optical & \\
Object & class$^\mathrm{a}$ & N & Comment & {\scriptsize MOJAVE$^\mathrm{b}$}\\
\noalign{\smallskip}
\hline
\noalign{\smallskip}
0048$-$097 & B & 5 & Compact, N emission        & Y     \\
0108$+$388 & G & 2 & EW structure       & N     \\
0119$+$115 & Q & 2 & N jet              & Y     \\
0201$+$113 & Q & 2 & NW jet             & N     \\
0218$+$357 & Q & 4 & Gravitational lens$^\mathrm{c}$ & N     \\
0235$+$164 & B & 5 & Compact, E emission        & Y     \\
0310$+$013 & Q & 1 & Compact, SE emission       & N     \\
0405$-$385$^\mathrm{d}$ & Q & 2 & W jet              & N     \\
0420$+$022 & Q & 2 & Compact    & N \\
0521$-$365$^\mathrm{d}$ & G & 3 & Long NW jet, difficult comp.\ id.\,& N     \\
0552$+$398$^\mathrm{d}$ & Q & 4 & Compact, W emission  & Y   \\
0602$+$673 & Q & 4 & Compact & N        \\
0615$+$820$^\mathrm{d,e}$ & Q & 7 & Compact, halo  & N    \\
0723$-$008 & B & 2 & NW jet             & N     \\
0808$+$019 & B & 4 & Compact, S emission & Y \\
0859$+$470 & Q & 2 & N jet              & N     \\
1032$-$199 & Q & 2 & SW emission        & N     \\
1124$-$186 & Q & 2 & Compact, S emission        & Y     \\
1155$+$251 & Q & 4 & Too complex & N    \\
1328$+$254 & Q & 4 & Too complex & N    \\
1328$+$307 & Q & 4 & Too complex & N    \\
1354$-$152 & Q & 2 & Compact, NE emission & N   \\
1424$+$366$^\mathrm{d}$ & Q & 5 & Compact, SW emission       & N     \\
1504$-$167 & Q & 2 & SE jet             & Y     \\
1511$-$100 & Q & 2 & E jet              & N     \\
1514$-$241 & B & 2 & S jet              & N     \\
1519$-$273 & B & 2 & Compact, W emission & N    \\
1739$+$522 & Q & 2 & Compact, NE emission       & Y     \\
1741$-$038 & Q & 2 & SW emission                & Y     \\
2155$-$152 & Q & 2 & SW jet             & Y     \\
2255$-$282 & Q & 2 & SW jet             & N     \\
\hline
\end{tabular}
}
\]
\begin{scriptsize}
\begin{list}{}{
\setlength{\leftmargin}{6pt}
\setlength{\rightmargin}{0pt}
}
\item[$^{\mathrm{a}}$] Optical classification according to the 
V\'eron-Cetty \& V\'eron (\cite{ver01}) catalog, where Q\,=\,quasar,
B\,=\,BL\,Lac object, 
and G\,=\,galaxy.
\item[$^{\mathrm{b}}$] Belonging to the complete sub-sample (MOJAVE, 
observed from May 2002 on).
\item[$^{\mathrm{c}}$] 
A detailed study
of the relative motions within the sub-images A and B and between them
will be presented elsewhere.
\item[$^{\mathrm{d}}$] Includes data from the pre-VSOP observations
by L.\ I.\ Gurvits et al.\ (in prep.).
\item[$^{\mathrm{e}}$] Includes data from the S5 
Polar Cap Sample (P\'erez-Torres et al.\ \cite{per04}).
\end{list}
\end{scriptsize}
\end{flushleft}
\end{table}

Apart from the 110 sources reported in Kellermann et al.\ (\cite{kel04})
and the 19 ones in this paper, some objects remain without 
kinematic analysis.  Those objects are listed in Table \ref{table:nokin}.
From the remaining objects, some of them have such a structure that 
prevents a reliable analysis (e.g., 0521$-$365 hosts 
a long jet and the low declination makes mapping and model fitting to 
be difficult).
A fraction of the objects have a parsec-scale jet where kinematics
can be measured, but only two observing epochs have been observed
so far.  Some of those\footnote{0119+115, 1504$-$167, 1741$-$038
(OT\,$-$068), and 2155$-$152 (OX\,$-$192).}  
are being monitored at the MOJAVE program, and therefore  reliable 
kinematics will be available
in the near future.  
The rest\footnote{0108+388 (OC\,+314), 0201+113 (OD\,+101), 
0405$-$385, 0723$-$008 (OJ\,+039), 
0859+470 (4C\,+74.29), 1511$-$100 (OR\,$-$118), 
1514$-$241 (AP\,Librae), and 2255$-$282.} are not being observed further,
so unless alternative
observations become available, no kinematic study using more than the
existing two
epochs will be possible.
Finally, the objects not categorised above are compact, and some 
detailed $(u,v)$-model fitting
can eventually  reveal changes in the structure, but 
higher frequency observations should provide the final answer
on their kinematics.

The complete set of kinematics
for the 110 sources reported in Kellermann et al.\ (\cite{kel04})
and the newer ones, together with an analysis of the component 
$(x,y)$ motions and the flux density evolution in selected cases 
will be reported in Ros et al.\ (in prep.).

%
\section{Compact, but slightly resolved sources\label{sec:indiv}}

We have reported in detail in Sect.~\ref{sec:kinematics} 
the kinematics of the sources
with a core-jet morphology.  In our sample we also find many
compact objects.
A careful look
to these objects may yield surprises, since not all of them
are unresolved. Here we show three typical cases of this kind.
An analysis of the fine scale structure based on the 
interferometric visibilities 
is presented in Y.\ Y.\ Kovalev et al.\ (in prep.).



\begin{figure}[t!]
\begin{center}
\includegraphics[clip,width=\columnwidth]{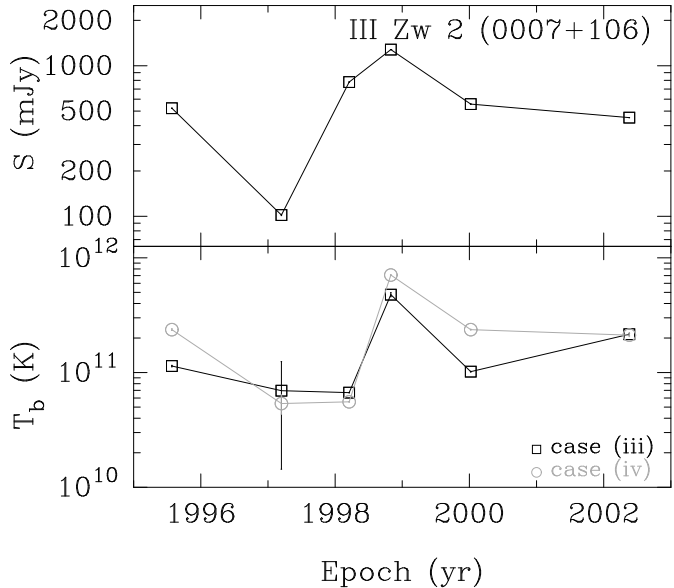}
\end{center}
\vspace{-15pt}
\caption{Time evolution of the total flux density and brightness temperature in III\,Zw\,2, 
derived from our model fitting (cases (iii) and (iv)).  The
error bars are formal standard errors.
\label{fig:iiizw2-tb}
}
\end{figure}

\paragraph{\textbf{III\,Zw\,2 (0007+106)---cracking the nut ?}}
The first panel in Fig.~\ref{fig:kinematics} reports the
proper motions 
III\,Zw\,2 derived from a model fitting
of the interferometric visibilities using two circular
Gaussian components---case (i); the use of two delta
distributions (case (ii)) yields equivalent results.
This Seyfert 1 galaxy, also
classified as ``radio intermediate quasar", is highly variable,
and a radio flare (Falcke et al.\ \cite{fal99})
happened in the course of our VLBI monitoring, at epoch $\sim$1998.8.
Brunthaler et al.\ (\cite{bru03})
analysed the structure of the radio source at 15\,GHz
model fitting the visibilities with two point components in nine epochs from
1998.2 to 2000.7.  Their data show a constant change in the separation of 
the components with speeds of $\sim0.6\,c$.  A direct comparison
of their results and our plot show discrepancies, since
our model at epoch 1998.21 shows a larger component separation.
although our 1998.83 and 2002.02 data points are compatible with
their measurements.
%
A further approach to
understand the changes in the source is to model fit its 
structure with a single circular (case (iii)) or elliptical
(axis ratio 0.4, case (iv)) Gaussian component.
The flux densities and the corresponding brightness temperatures
are shown in Fig.~\ref{fig:iiizw2-tb}.  
Notice that the brightness temperature was stable  while the flux
density was rising, and remained low at the highest state
of the flare, becoming much higher in the post-flare phase
and coming slowly back to the values of the rising phase
at the last epoch.
We can compare this result with the suggestions of 
Brunthaler et al.\ (\cite{bru03}) that the source
can be explained with an ``expanding balloon" model, say,
``a radio galaxy in a nutshell".
Our comparatively low brightness density for the rising and peaking
phase of the flare is in apparent contradiction with the
interpretation of the 15\,GHz data being the post-shock material
from the hot-spots expanding in
ultra-compact scales and interacting with the environment.
In any case, we observe the same process of inflation and deflation
in the source, but with a sparser time sampling and at only one frequency.

\paragraph{\textbf{AO\,0235+16 (0235+164) \& NRAO\,512 (1638+388)}}

\begin{figure*}[t!]
\begin{center}
\includegraphics[clip,width=0.9\textwidth]{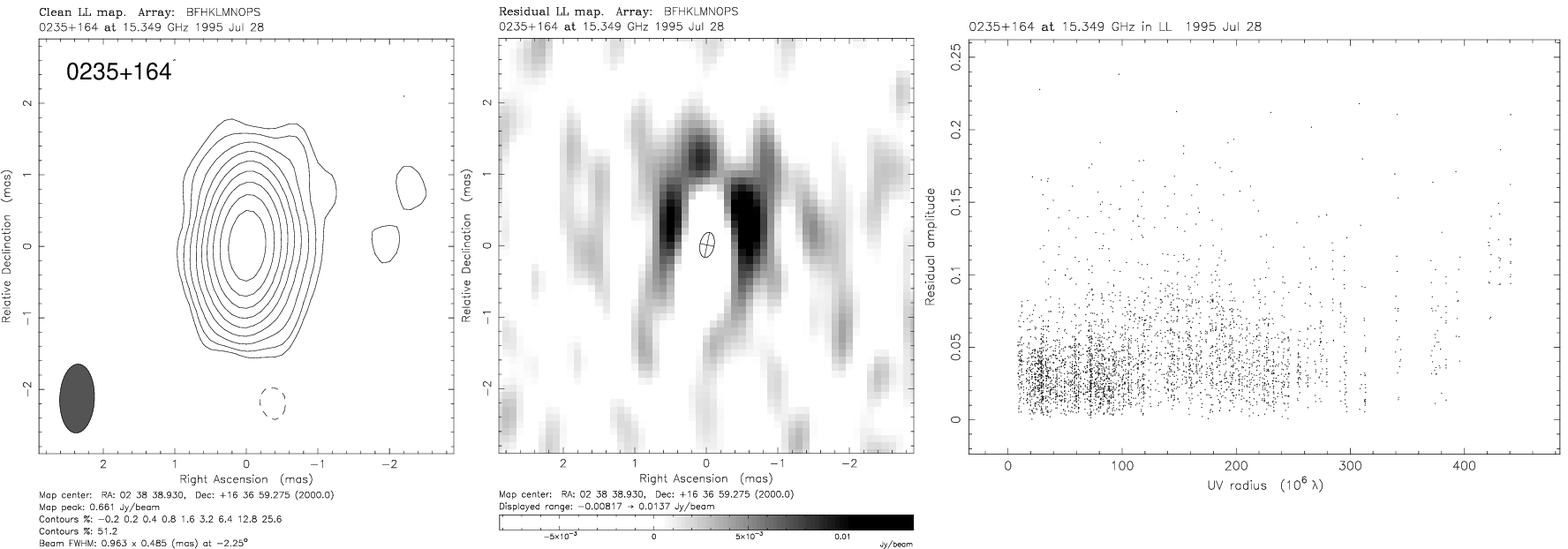}
\vspace{10pt}
\includegraphics[clip,width=0.9\textwidth]{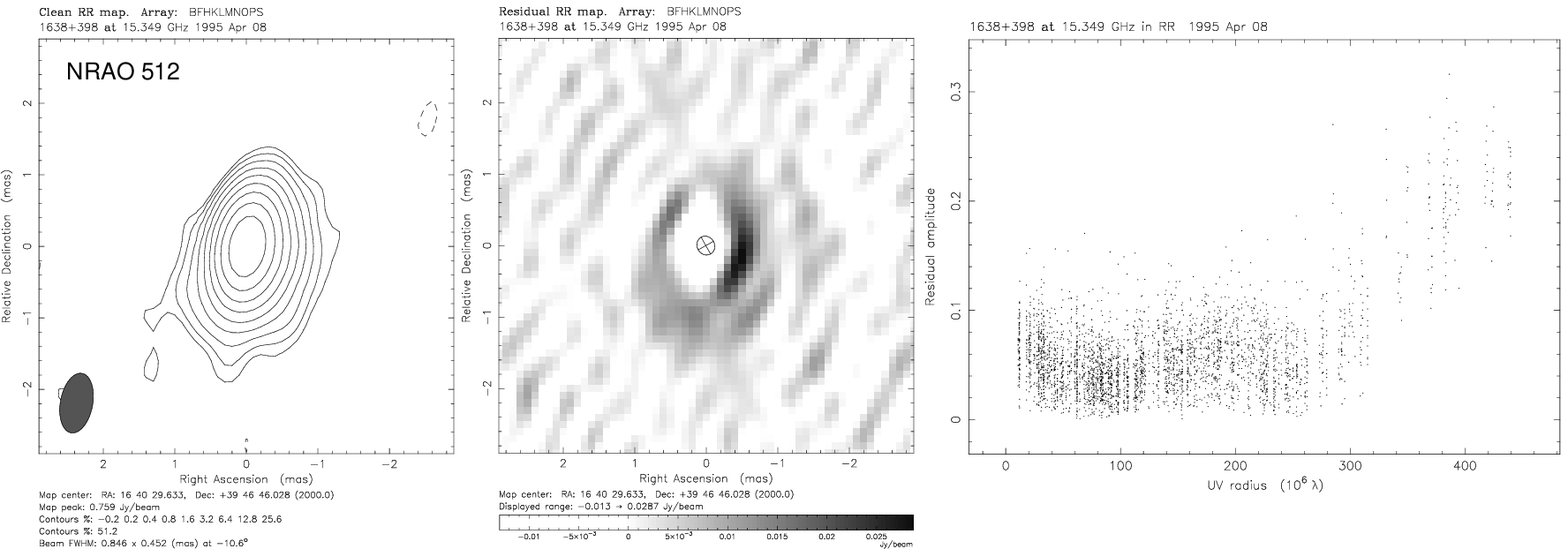}
\end{center}
\vspace{-20pt}
\caption{Images and visibility residuals from the compact,
but resolved sources AO\,0235+16 and NRAO\,512 at their first
observing epochs in the 2\,cm Survey after fitting them with
a single Gaussian component.  Left panels:
hybrid maps.  Central panels: residual maps after fitting a single
elliptical Gaussian component to the visibilities (with flux [mJy],
major axis [mas], axis ratio, and position angle [deg], respectively
of 766:0.357:0.55:$-11.4$ and 914:0.271:0.924:24.0).  The profile
of the ellipse is shown at the centre, being the two major axes
the full-width half maxima.  Notice the need for extended emission
beyond one beam size, shown in gray scale.
Right panels: residuals of the visibility amplitude in Jy versus
the interferometric distance, showing
the result of subtracting the Gaussian model fit from the observed
data.
\label{fig:0235nrao512}
}
\end{figure*}

Those two 
AGN 
show compact structure (left panel in Fig.\ \ref{fig:0235nrao512})
and flux density variability. 
The BL\,Lac object AO\,0235+16 shows intra-day variability 
(Quirrenbach et al.\ \cite{qui92}; Kraus et al.\ \cite{kra99})
and recurrent outbursts with a period of $\sim$$(5.7\pm0.5)$\,yr 
(Raitieri et al.\ \cite{rai01}).  It is well known that
a very high brightness temperature
component is present at milliarcsecond-scales (e.g., 
Frey et al.\ \cite{fre00};
Peng \& de Bruyn \cite{pen04}).
The quasar NRAO\,512 shows unresolved structure at 8.4\,GHz or 2.3\,GHz (e.g., Fey
\& Charlot \cite{fey00}).  For this reason, it
has been traditionally used as a phase-reference
calibrator for the neighbour source 3C\,345 
(e.g., Bartel et al.\ \cite{bar86}). 
The {\sc clean} images shown at the 
left panels of Fig.\ \ref{fig:0235nrao512} shows that 
both NRAO\,512 and AO\,0235+16 are compact objects
but a detailed analysis shows that those are not unresolved.  
Figure \ref{fig:kinematics} shows measured 
kinematics for NRAO\,512 ussing a three-Gaussian model.  
The multi-epoch model fitting with three components appears consistent.

A rough analysis, presented in
Figure \ref{fig:0235nrao512}, shows that a single Gaussian fit to
the structure of those two objects does not adequately reproduce 
their structure.  The mid panels show the residual maps after this fitting.
The right panels show the residuals of this fit, and the discrepancy beyond
300\,M$\lambda$ is obvious, especially in the case of NRAO\,512.
A model with a clean component and an extended structure can better
fit the data---a sort of ``prussian hat".  An analysis of the 
change of the flux densities
of both the ultra-compact and the extended component over time will
be presented elsewhere.  As in other examples, 
the ultra-compact component is needed to explain
the intra-day variability (either if intrinsic or if caused by
inter-stellar scattering), and the extended emission is related
to a jet structure, that can be more or less weak, 
and face-on or slightly tilted and therefore present in the images.

The analysis of those sources and the rest of the 2\,cm Survey
will keep providing interesting results and shedding more light
on the nature of the AGN in radio frequencies and its relationship
with their properties in other wavebands (see, e.g., Kadler
et al., these proceedings, concerning X-ray observations).

%
%

\begin{acknowledgements}
This work has been made in collaboration with the 2\,cm\,Survey Team,
see {\tt http://www.nrao.edu/2cmsurvey}.
E.\,R.\ acknowledges the support
of the European Commission's I3 Programme
``RADIONET", under contract No.\ 505818 to attend the conference.
{The VLBA is operated by the National Radio Astronomy 
Observatory which is a facility of the National Science Foundation 
operated under cooperative agreement by Associated Universities Inc.}
\end{acknowledgements}

\end{document}